\documentclass[12pt,preprint]{aastex}
\shorttitle{Cluster Neutrinos}
\shortauthors{Wolfe}

\begin{document}

\title{Neutrinos and Gamma Rays from Galaxy Clusters}

\author{Brandon Wolfe\altaffilmark{1,5}, Fulvio Melia\altaffilmark{1,2}, Roland M. Crocker\altaffilmark{3,4},
and Raymond R. Volkas\altaffilmark{5}}
\altaffiltext{1}{Physics Department, The University of Arizona, Tucson, AZ 85721}
\altaffiltext{2}{Steward Observatory, The University of Arizona, Tucson, AZ 85721}
\altaffiltext{3}{School of Chemistry and Physics, The University of Adelaide, South Australia 5005}
\altaffiltext{4}{Current address: J.L. William Fellow, School of Physics, Monash University, 
Victoria 3800, Australia}
\altaffiltext{5}{School of Physics, Research Centre for High Energy Physics, University of 
Melbourne, Victoria 3010, Australia}

\begin{abstract}
The next generation of neutrino and $\gamma$-ray detectors should provide new insights
into the creation and propagation of high-energy protons within galaxy clusters,
probing both the particle physics of cosmic rays interacting with the background medium
and the mechanisms for high-energy particle production within the cluster.
In this paper we examine the possible detection of $\gamma$-rays (via the GLAST satellite)
and neutrinos (via the ICECUBE and Auger experiments) from the Coma cluster of galaxies,
as well as for the $\gamma$-ray bright clusters Abell 85, 1758, and 1914. These three
were selected from their possible association with unidentified EGRET sources, so it 
is not yet entirely certain that their $\gamma$-rays are indeed produced diffusively within
the intracluster medium, as opposed to AGNs. It is not obvious why these inconspicuous
Abell-clusters should be the first to be seen in $\gamma$-rays, but a possible reason
is that all of them show direct evidence of recent or ongoing mergers. Their identification
with the EGRET $\gamma$-ray sources is also supported by the close correlation between
their radio and (purported) $\gamma$-ray fluxes. Under favorable conditions (including
a proton spectral index of $2.5$ in the case of Abell 85, and $\sim 2.3$ for Coma,
and Abell 1758 and 1914), we expect 
ICECUBE to make as many as $0.3$ neutrino detections per year from the Coma cluster of 
galaxies, and as many as a few per year from the Abell clusters 85, 1758, and 1914.
Also, Auger may detect as many as 2 events per decade at $\sim 10^{18}$ eV from
these gamma-ray bright clusters. 
\end{abstract}


\section{Introduction}

A population of high-energy cosmic-ray (CR) protons, pervading galaxy clusters, was 
first invoked (Dennisson 1980) to explain highly polarized radio luminosities ($L_r 
\sim 10^{40}-10^{42}$ erg s$^{-1}$) now observed in over 30 clusters. Evidently the 
radio light is synchrotron emission from highly relativistic electrons ($\gamma_e 
\sim 10^6$) gyrating in large magnetic fields ($B \sim 1 \ \mu$G), despite the fact 
that such electrons would lose their energy to radiation well before they could 
traverse the $\sim$Mpc extent of observed emission. However, protons injected into 
the cluster diffuse throughout the intracluster medium, essentially unchanged and 
confined for cosmological times. In this environment they may interact with the 
background gas and create a decay cascade. Among the products of this decay are
the relativistic electrons responsible for radio emission.

Since Dennisson's (1980) pioneering work, conclusive observations of this `secondary' 
process for electron production within clusters have been few. In Hercules A (Nulsen 
et al. 2005b), MS 0735 (McNamara et al. 2005), and Hydra A (Nulsen et al. 2005b),
vacated areas of X-ray bremsstrahlung indicate the background of thermal electrons 
is being pushed aside by relativistic electrons injected from the axes of a central 
source, implying a phenomenally large power, $L_p \sim 10^{44}$ erg s$^{-1}$. These 
`bubbles' coincide with sources of radio emission as the protons produce a cascade 
of synchrotron-emitting electrons, confirming that the secondary decay process does 
occur in clusters. The protons released during such an episode would be confined and 
built up within the cluster over cosmic times, totaling some $10^{61}$ erg of nonthermal 
hadrons---very near energy equipartition with the thermal background (Berezinsky, Blasi, 
and Ptuskin 1997; Hinton \& Domainko 2006). This is, we find, the order of nonthermal
energy required to be deposited in the nonthermal hadron population for proton-proton
scatterings to produce $\pi^0$ decay $\gamma$-rays within the EGRET detection sensitivity.

The energy liberated by the infall of the galaxy group NGC~4839 toward the center of 
the Coma cluster, observed by XMM-Newton (Neumann et al. 2001), is an example of another 
potential source of high-energy protons. NGC~4839 achieves a velocity of $\sim 1,400$ km 
s$^{-1}$, and since the sound speed corresponding to Coma's gas temperature of $\sim 8$ 
keV is $\sim 1,000$ km s$^{-1}$, the subcluster's supersonic motion is expected to 
produce shocks (which Neumann et al. claim to observe directly in the imaging of 
NGC~4839). It is likely that the shock compression in these regions energizes a fraction
of the thermal particles within the intracluster medium (ICM) by first-order Fermi
acceleration. However, how the protons are actually energized is not directly relevant
to our analysis. For example, since the cosmic rays remain trapped within the cluster 
for over a Hubble time, it is likely that second-order acceleration by a turbulent 
distribution of Alfv\'en waves may be even more effective at helping the protons attain 
their highest energies (see, e.g., Liu, Petrosian, and Melia 2004; Liu, Melia, and
Petrosian 2006). Our principal goal is to determine the observational signatures 
such a population would produce, should it be present within the cluster. But cluster 
mergers do provide us with a viable mechanism for energizing the cosmic rays, regardless 
of how the individual particles are ultimately accelerated. In this regard, we note that
the gravitational energy $\sim 10^{64}$ erg released in a large relic merger could 
reasonably supply $\sim 10^{62}$ erg in a nonthermal hadron population, again 
sufficient to supply detectable $\gamma$-rays. In this case, the radio halo may 
be interpreted as a signature of the cosmological development of a cluster (the 
oldest known cluster being some 9 billion years old).

Aside from a radio halo produced by secondary electrons,
conclusive evidence for the properties of CR protons within galaxy clusters
rests on the detection of other products of the protons' decay cascade: $\gamma$-rays
and neutrinos. The closest active galaxy cluster, the Coma cluster, does not emit
copiously enough to have been detected by EGRET given its $\sim10^{-11}$ cm$^{-2}$
s$^{-1}$ flux sensitivity, although we will show that this does not
preclude a possible detection by the GLAST satellite or a
detectable neutrino flux. Our preliminary identification of clusters which \emph{are}
$\gamma$-ray bright, and, therefore, plausible extrasolar neutrino candidates,
has come from a systematic search for likely clusters among unidentified EGRET sources.

In a search for new radio galaxy clusters, the NRAO VLA (Giovannini et al. 1999) and
Westerbork Northern (Kempner \& Sarazin 2001) sky surveys were systematically cross
referenced with X-ray bright objects from the ROSAT survey (Ebeling et al.
1998), providing a list of more than 30 known cluster halos or relics,
presented in Table 1. While none of the clusters EGRET was hoped to detect
produce a measurable $\gamma$-ray flux (Reimer et al. 2003), three radio
sources coincide with unidentified $\gamma$-ray bright objects detected by EGRET
(Colafrancesco 2002)---the cluster Abell 85 and the EGRET source 3EG J0038-0949,
Abell 1914 and EGRET source J1424+3734, and Abell 1758 and EGRET source 3EG J1337 +502a.
These three EGRET sources are not unequivocally the quoted clusters, but here we
shall assume they are in fact the same objects.

The assumption is not certain. In a study by Reimer et al. (2003) it was shown that
the probability that one of the 170 unidentified EGRET sources and a member of
their list of 58 considered galaxy clusters coincided was 48.1\%.
Of their list of 58 clusters worthy of consideration, only
Abell 85 and the unidentified source J0038-0949 showed a considerable overlap,
which could therefore be explained as an entirely chance occurrence.
However only six of their 58 considered clusters
---included for their proximity ($z<0.14$) and X-ray brightness---
were shown by the NRAO VLA or Westerbork Northern
radio sky surveys to be radio halo/relic clusters, so that in only these six
cases is there any reason to believe a significant nonthermal proton population may be present.
Also, Abell 1758 and 1914 were not considered by Reimer et al. (2003), as they are above $z >0.14$.

The three candidates are not the most X-ray bright clusters, most of which were 
assigned only upper limits in the nine years of EGRET observations. It is not 
obvious why three inconspicuous Abell-clusters are the first to be seen in gamma 
rays when the closest and most well-studied clusters---known to have evidence for 
non-thermal particle populations from radio, EUV, or hard X-ray detections---are not.
Perhaps a key is that
each of the $\gamma$-ray bright candidates shows direct evidence of recent or
ongoing mergers, indicating (if confirmed) that this may outcompete the AGN outburst
as a source of cluster cosmic rays. In Abell 85, a 4 Mpc filament (Durret
et al. 2005) forms a chain of several groups of galaxies.
In particular, an impact region (the `south blob') displays a temperature ($8.8$ keV) far
in excess of the rest of the cluster ($2.7$ keV). In Abell 1914 (Govoni et al. 2004)
a NE-SW arch-like hot region extends across the cluster
center and may represent an ongoing shock. Abell 1758 is a double cluster
(David \& Kempner 2004): A1758N is in the late stages of a large impact
parameter merger between two $7$ keV clusters, while A1758S is
undergoing a smaller impact parameter merger between two 5 keV cores.
In each case, the flux of $\gamma$-rays correlates with the radio flux, as it would if
both were linked to secondary decay products. Abell 85 also shows evidence of a
hard X-ray excess (Lima-Neto et al. 2001), indicative of
a nonthermal population of electrons. 

In the end, however, we caution that while the identification of Abell clusters 
85, 1758, and 1914 with three unidentified EGRET sources may be motivated by the 
reasons we have outlined here, it is not entirely clear that their $\gamma$-rays 
are indeed produced diffusively within the ICM. Thus, one must accept the results 
presented in this paper, particularly the predicted neutrino fluxes, with this 
important caveat in mind.

As clusters confine CR protons over cosmological times, they provide a promising
source for a possible first detection of high-energy ($\gtrsim$ TeV) extra-solar
neutrinos. Aside from probing high or ultra-high energy particle physics processes 
occurring within astrophysical sources, any such neutrino (or $\gamma$-ray) detection
would also establish the properties of the proton population within the cluster
and settle the possible role of such populations in various cluster
phenomena---e.g., generating the radio halo (Dennisson 1980), their contribution to
the `cooling flow' problem (Silk 1995), or their acceleration within cluster
merger shocks (Sarazin 2004).

In this paper we take the optimistic view that protons within the Coma cluster
create a flux of $\gamma$-rays which is within the detection limits
of the GLAST satellite, without violating the EGRET null detection.
In the same spirit of optimism, we identify the Abell clusters 85, 1758, and 1914 with
their gamma-ray bright counterparts. The resulting predictions for neutrino
and gamma-ray detections for these four clusters will determine whether this optimism was
well-founded.

Work on the hadronic origin of one or more spectral components in the overall
cluster emissivity has been quite extensive, and our investigation overlaps with several 
earlier treatments, though our model is unique in the treatment of the particle
kinetics and proton-electron scattering events. Dolag and Ensslin (2000) considered
the possible origin of the radio halos from a hadronic secondary electron injection,
based on a detailed modeling of the cluster magnetic field (see also Blasi \& 
Colafrancesco 2000, Jones 2004, Brunetti et al. 2007, Marchegiani et al. 2007). 
Subsequently, Pfrommer and Ensslin (2004) constrained the cosmic-ray population 
in several cooling flow clusters from their radio and $\gamma$-ray emissivities (see 
also Reimer et al. 2004, and Sarazin 2007). Not surprisingly, the $\gamma$-ray emission 
from proton-proton collisions in clusters may contribute a non-negligible fraction
of the extragalactic background radiation field (Kuo et al. 2005). And though
we will not be addressing $\gamma$-ray line emission in this paper, cosmic rays may
also collide with heavier nuclei in the ICM to produce measurable $\gamma$-ray
line intensities (Iyudin et al. 2004). An extensive review of the $\gamma$-ray 
emissivity in clusters may be found in Blasi et al. (2007). 

The issue of hadronic acceleration during large-scale structure formation has been 
considered by several authors, including Miniati et al. (2001), Berrington \& Dermer 
(2003), Blasi (2004), Ensslin et al. (2007), Jubelgas et al. (2008), Ando and Nagai 
(2008), Nakar et al. (2008) and, in the context of secondary acceleration due to 
stochastic processes, Brunetti et al. (2004), Brunetti and Blasi (2005), and Petrosian 
and Bykov (2008). But cosmic rays may also play an important dynamical role throughout 
the cluster's lifetime. Collisions between energetic hadrons and the ICM have been 
invoked as a means of heating the gas condensing toward the middle of the cluster, 
thereby preventing an overly rapid cooling of the flow (Colafrancesco, Dar, and 
De R\'ujula 2004, Guo \& Oh 2008). 

Our goal in this paper is specifically to calculate the neutrino flux expected
from $\gamma$-ray bright clusters, but our approach differs from earlier attempts
in several important ways, which we will describe over the next several sections. 
Our model requires that the number of protons 
be consistent with a secondary explanation of the Coma cluster's radio halo, using 
magnetic fields ($\sim 1 \mu$G) consistent with observations of Faraday rotation, 
rather than those ($\sim 0.1 \mu$G) consistent with an inverse-Compton description 
of the hard X-ray excess detected by Beppo\emph{SAX} and RXTE (see also Wolfe and 
Melia 2006). The only model-dependent characteristics of this work are then the
distribution of gas and that of protons diffusing throughout the cluster, which---together
with the details of calculating the various links in the decay cascade---we discuss in 
Section 2. Our technique for calculating the detected number of neutrinos, given the predicted flux
of Coma, is discussed in Section 3. Finally in Section 4, we review the forecast for
those clusters for which a positive $\gamma$-ray detection has been proposed.

\section{Model Characteristics and Flux Calculation}
\subsection{Emission spectra of nonthermal plasmas}
Our model first requires the distribution of electrons, neutrinos, and $\gamma$-rays
produced by proton-proton collisions and subsequent decay of secondaries.
Gamma rays are produced from the decay of neutral
pions $pp \rightarrow \pi^0 \rightarrow 2 \gamma$ at a
characteristic energy of $\sim$ 70 MeV in the pion rest frame;
a similar number of neutrinos are also produced via
$pp \rightarrow \pi^\pm \rightarrow \mu^\pm + \nu_\mu(\bar{\nu}_\mu)$,
and in the subsequent decay of muons into electrons and positrons, $\mu^\pm \rightarrow
e^\pm + \bar{\nu}_\mu (\nu_\mu) + \nu_e (\bar{\nu}_e)$ (given
typical power-law parent proton distributions,
neutrinos from neutron $\beta$-decay
may be neglected in these calculations). Both neutrinos and
$\gamma$-rays may also be produced via proton interactions with background
light fields. For the cosmic microwave background, pion production is characterized
by a peak at $6 \times 10^{19}$ eV---approximately the energy of a 140 km/h tennis
ball---while for infrared light emitted by galaxies and proposed
(Blasi 2005) as a dominant source, this peak shifts to $2.5 \times 10^{18}$ eV.
All three decay sources probe different energy regimes: electrons, via synchrotron radio
($10^{-6}$ eV) or HXR inverse-Compton ($10$ eV); $\gamma$-rays, at 70 MeV
and notably detectable by GLAST and EGRET; and TeV+ neutrinos, which directly probe the
tail of the proton distribution ($10^{13}$--$10^{15}$ eV) via pp scatterings as well
as its extreme limit ($>10^{18}$ eV) via interactions with the CMB.

Our recipe is to assume or imply (via radio observations, see Sec 3.2) a proton
distribution, which in a practical sense means setting a spectral index and
normalization to a power-law $n_p(E_p) = n_{p0} E_p^{s_p}$. From this alone follow
two injections of pions---one for proton interactions (pp) with hydrogen in
the ICM, and one for interactions with background light (p$\gamma$). The
observationally relevant decay products of these pions are electrons,
$\gamma$-rays, and neutrinos (see also Blasi and Colafrancesco 1998), which 
for the particles of interest may be
found by summing over the kinetically allowed energies in each link of a
decay chain; this section is devoted to a description of this process.

For proton-proton interactions, the neutral pion emissivity is
\begin{equation}
q^{pp}_{\pi^0}(E_{\pi^0}) = c n_H \int_{E_{thres}} dE_p n_p(E_p) \frac{d \sigma(E_{\pi^0},E_p)}{d E_{\pi^0}}\;,
\end{equation}
and similarly for charged pions, via an altered cross section.
The threshold proton kinetic energy for production of a pion in a proton-proton collision
is $E_{thres} = 2 m_\pi (1 + m_\pi/4 m_p)$.

Here we take the cross-section
for pion production for energies below the isobar resonance ($E_p = 7$ GeV) from
Dermer (1986)---where also each of the various decay channels is treated---while
above the resonance we take the cross-section to be a simple scaling form
used by Blasi $\&$ Colafrancesco (1999), and Fatuzzo $\&$ Melia (2003).
We note in this context that while collider data at these extreme energies
do not exist, the proton-proton collision cross-section has
been examined by the Fly's Eye detector, which implies from atmospheric
fluorescence that the collision cross-section violates the
scaling assumption when $\sqrt{s}$ is tens of TeV.
Unfortunately the pion production cross-section is determined both
by the rate of proton collisions, $R_{pp} = n_H c\, d\sigma_{pp}/dE_p$,
and by the varying multiplicity, $M(E_\pi,E_p)$, of pion production with proton energy
(Moffeit et al. 1972, Markoff and Melia 1997, 1999). For this we again
depend on accelerator data, which exist only up to 1,800 GeV---so that the
limitation on our knowledge of the cross-section is compounded by a lack of
certainty regarding pion production. Simply, predictions and observations of
neutrinos or $\gamma$-rays in clusters probe energy regimes not previously accessible.

Once the pion source function is found, the neutrino, $\gamma$-ray, and electron fluxes
follow from kinematic concerns discussed fully in Stecker (1979) for $\gamma$-ray decays
and Marscher et al. (1980), and Zatsepin $\&$ Kuz'min (1962) for neutrinos.
For decay photons we have
\begin{equation}
q^{pp}_{\gamma}(E_\gamma) = 2 \int_\phi dE_{\pi^0} \frac{ q^{pp}_{\pi^0}(E_{\pi^0}) }
	{(E_{\pi^0}^2 - m_{\pi^0}^2 c^4)^{1/2}}\;,
\end{equation}
where the minimum kinetically allowed pion energy is $\phi = E_\gamma + m^2_{\pi^0}
c^4/(4 E_\gamma)$. Well above the 70 MeV region,
the pion spectrum shares the proton parents' spectral index, $s_\pi = s_p$.
For electrons, the source function is
\begin{equation}
q^{pp}_e(E_e) = n_H c \frac{m_\pi^2}{m_\pi^2 - m_\mu^2}
\int^{E_p^{max}}_{E_e}
dE_\mu \frac{dP}{dE_e}
\int^{E_\pi^{max}}_{E_\pi^{min}}
\frac{dE_\pi}{\beta_\pi E_\pi}
\int^{E_p^{max}}_{E_{thres}(E_\pi)}
dE_p n_p(E_p) \frac{d\sigma(E_\pi,E_p)}{dE_\pi}\;,
\end{equation}
where $dP/dE_e$ is the three-body decay probability (Fatuzzo \& Melia 2003). Here, the limits
of kinematically allowed pion energies are
\begin{equation}
E_{\pi}^{min,max} = \frac{2 E_\mu}{(1\pm\beta_\mu)+m_\pi^2/m_\mu^2(1\mp\beta_\mu)}
\end{equation}
(with minimum corresponding to the upper sign, maximum to the lower).
The threshold $E_{thresh}(E_\pi)$for producing a pion with energy $E_\pi$ is similar to that in Eq. (1).

For neutrinos, we must distinguish between particles produced by pion decay, and those
produced by muon decay, since these are two- and three-body decays, respectively.
For pions the source function is
\begin{equation}
q^{pp}_{\nu_\alpha}(E_{\nu_\alpha}) = \int_{\gamma^\pi_{min}}^\infty
	f(\gamma_\pi) q_\pi(\gamma_\pi) d\gamma_\pi\;.
\end{equation}
In the case of electron injection resulting from proton collisions we integrated over the three-body
decay probability $dP/dE_e$. Here $f(\gamma_\pi)$ plays that role, as the spectrum of two-body decay
\begin{equation}
f(\gamma_\pi) = \frac{1}{2 E_\nu^0 \sqrt{\gamma_\pi^2 - 1}}\;.
\end{equation}
Here, $E_\nu^0 \approx 29.8$ MeV is the CMS energy of the emitted neutrino. In this
instance the minimum kinetically allowed energy is
\begin{equation}
\gamma_{min}^\pi = E_\nu^0/2E_\nu + E_\nu/2E_\nu^0\;.
\end{equation}

In the decay of charged muons, the muon neutrino flux is
\begin{equation}
q_\nu^\mu (E_\nu) = \frac{1}{2} \int_{\gamma_\mu,min}^\infty f(E_\nu) \int^1_{\chi_{min}} q_\pi(\gamma_\pi)
	\frac{\partial \gamma_\pi}{\partial \gamma_\mu}\; d\chi \;d\gamma_\mu\;,
\end{equation}
where electron neutrinos are produced at approximately half the rate of these muon neutrinos.
The interaction angle cosine, $\chi$, is kinematically limited to
\begin{eqnarray}
\chi_{min}&=&-1\,,\quad{\rm for}\;  3.68 \le 3.55 \gamma_\mu\nonumber\\
      &=&\sqrt{13.54 - 12.6 \gamma_\mu^2}\,,\quad{\rm for}\;  3.68 > 3.55 \gamma_\mu\;,
\end{eqnarray}
while the lower limit for $\gamma_\mu$ is
\begin{equation}
\gamma_{\mu,min} = {E_\nu\over m_\mu} + {m_\mu\over 4 E_\nu}\;.
\end{equation}
In this case we require $f(E_\nu)$, the spectrum of three-body decay, which is
\begin{eqnarray}
f(E_\nu)&=&16 \gamma_\mu^5 \bigg( {3\over\gamma_\mu^5} - {4\over 3}(3+\beta_\mu^2)\zeta \bigg) \zeta^2 {1\over m_\mu}\,,\;{\rm for}\; 0 \le \zeta \le [1-\beta_\mu]/2\nonumber\\
&=&{ {5\over 3} + {4\over (1+\beta_\mu)^3} \bigg( {8 \zeta\over 3} - 3(1+\beta_\mu) \bigg) \zeta^2}{ \beta_\mu \gamma_\mu m_\mu }\,,\;{\rm for}\; [1-\beta_\mu]/2 \le \zeta \le [1+\beta_\mu]/2\;,\qquad
\end{eqnarray}
where $\zeta = ({E_\nu\over\gamma_\mu m_\mu})$ and $\beta_\mu = v_\mu/c$.
Since detection lies far above 70 MeV,
it suffices to approximate neutrinos
from pion decay as the same in number as those from subsequent muon decay.

In the case of proton-photon interactions, we may use the expressions (2, 5, \&8) unaltered,
changing only the source function of pion production. This is now given by
(Bottcher \& Dermer 1999)
\begin{equation}
q_\pi (\gamma_\pi) = {c\over 2} \int_1^\infty d\gamma_p\, n_p(\gamma_p) \int^1_{-1} d\chi
			\int_0^\infty d\epsilon\; n_\gamma(\epsilon) (1-\beta_p \chi) {d^2 \sigma\over d\gamma_\pi d \chi}\;,
\end{equation}
where $n_p$ and $n_{\gamma}$ are, respectively, the proton and photon distributions.
The result is a flat neutrino emissivity
$dN_\nu/dE\;dV\;dt$ until threshold, with a spectral index one higher than that of the proton
distribution afterwards, $s_\nu = s_p + 1$. Again, $\chi$ is the interaction angle cosine.

The cross-section for charged photopion production (Mucke et al. 2000, and references
cited therein) features two prominent decay resonances (Fig. 1): the first, at $E_\Delta = 330$ MeV,
has a height of $2.45 \times 10^{-28}$ cm$^2$ and a decay width of 0.15 GeV. The second
resonance, at 710 MeV, has a decay width of 0.18 GeV---but, due to the steepness of the proton
distribution, it does not play a major role in the neutrino flux. Note that this height is
some factor of 1.5 greater than that assumed by Dermer and Bottcher (2003).
The cross-section for neutral photopion production shows only a single resonance
of height $3 \times 10^{-28}$ cm$^2$ and a decay width of 0.15 GeV. Our fitting function
for a resonance at mass M, of height $\sigma_0$ and of width $\Gamma$ is of Breit-Wigner form,
\begin{equation}
f(\epsilon^\prime) = {\sigma_0 \Gamma^2 \epsilon^\prime\over(\epsilon^\prime-M^2)^2 + \Gamma^2 \epsilon^\prime}\;,
\end{equation}
where $\epsilon^\prime = \gamma_p \epsilon (1-\mu)$ is the photon energy in the proton rest frame.
We have tested the calculation with a simple $\delta$-function resonance, however, and the
results are nearly identical.

Finally, we require the synchrotron emission from injected electrons to set the level of
ambient relativistic protons. In practice this is evaluated numerically, but a sufficient
guide may be given by the approximation for the electron injection (Mannheim \& Schlickeiser 1994),
\begin{eqnarray}
q_e&=& {m_\pi\over 70\, m_e} q_{\pi^\pm} \bigg({E_\pi\over 70\;\hbox{MeV}}\bigg)
\sim {13\over 12}\sigma_{pp}\, c\, n_{H}\, n_{p0}(r) \bigg({m_p\over 24\, m_e} \bigg)^{s_{e0}-1}
(\gamma_e \beta_e)^{-s_{e0}}\;\; {\hbox{cm}}^{-3}\; {\hbox{s}}^{-1}\nonumber \\
            \null&\equiv&K_{inj}\, {\gamma_e}^{-s_{e0}}\;.
\end{eqnarray}
Assuming the dominant loss mechanism is either inverse Compton scattering with a blackbody
photon background ($T_{CMB} = 2.73$ K), or radio synchrotron, radiative losses are given
by $-dE_e/dt = a_s E_e^2$, with the constant $a_s = (4/3) \sigma_T\, c\, n_e (\epsilon_{CMB}
+ \epsilon_{B})/m_e c^2$. The energy density ($\epsilon_{CMB}$) in the CMB dominates
over that ($\epsilon_B$) in the magnetic field.  The equilibrium distribution of
electrons due to injection against these losses is (Blumenthal and Gould 1970)
\begin{equation}
n(\gamma_e) = {K_{inj}\over m_e c^2 a_s (s_{e0}-1)}\; \gamma_e^{-(s_{e0}+1)} \;.
\end{equation}
The radio synchrotron emissivity (in units of energy per unit volume, per unit time,
per unit frequency) associated with this distribution is then
\begin{equation}
{dE\over dV\, d\nu\, dt} \approx 1.15\, \pi^2 {K_{inj}\, \alpha\, \hbar\,
\nu_B\over m_e c^2\, a_s (s_{e0}-1)} \bigg({\nu_B\over\nu} \bigg)^{s_{e0}/2}\;,
\end{equation}
where $\nu_B$ is the gyrofrequency. For a final handle on the electron
population in clusters based on the presence or absence of a hard X-ray excess,
the corresponding Compton scattering emissivity
off the CMB (in units of photon number per unit volume, per unit time, per unit energy) is
\begin{equation}
{dN_\gamma\over dV\, d\epsilon\,dt} = 1.8{r_0^2\over\hbar^3 c^2}{K_{inj}\over m_e c^2 a_s
(s_{e0}-1)} (kT_{CMB})^{(s_{e0}+6)/2} \epsilon^{-(s_{e0}+2)/2}\;,
\end{equation}
where $r_0$ is the classical electron radius, and we have used the fact that the hard
X-radiation is produced below the Klein-Nishina region to simplify the cross section.

\subsection{Cluster modeling}
We now have spectra or distributions for each of the links in the decay cascade---electrons,
$\gamma$-rays, and neutrinos---and have derived expressions relating electron number to observed radio or nonthermal X-ray emission. It remains only
to assume a radial distribution for the background gases and for energetic protons
diffusing throughout the cluster. For simplicity we will adopt a single
point source of protons at the center of the cluster:
realistic morphology, time dependence, or non-Kolmogorov proton diffusion
through the cluster may seem to increase the accuracy of the model, but without a motivating
observation only serve to make it less reproducible. Below we describe in detail
how we calculate the $\gamma$-ray and neutrino fluxes from Coma, the results
of which are presented graphically in Fig. 2.

We take the background gas within Coma to be described by a $\beta$ model
(Colless \& Dunn 1996) with $n_{core} = 3 \times 10^{-3}$ cm$^{-3}$,
$\beta = 0.7$, and $r_{core} = 0.25$ Mpc, for which
\begin{equation}
n(r) = n_c \bigg[ 1+ \bigg({r\over r_c} \bigg)^2 \bigg]^{-3\beta/2}\;.
\end{equation}
While the thermal and nonthermal electron populations are well determined
by X-ray and radio observations (Fig. 3), respectively, no similar tool directly
determines the nonthermal proton population. Here, we assume high-energy
proton injection takes place at a central source and these then diffuse
outward according to (Blasi \& Colafrancesco 1999)
\begin{equation}
{\partial n_p(E_p,r,t)\over\partial t} = D(E_p) \bigtriangledown^2 n_p(E_p,r,t) + Q(E_p) \delta(r) +
{\partial\over\partial E_p} b(E_p) n_p (E_p,r,t)\;.
\end{equation}
The highest possible energy attainable by protons is actually given (Dermer \& Berrington 2002)
by the duration of acceleration, rather than any loss term $b(E_p)$.
Loss, and an additional term representing proton escape from the cluster, we neglect.
Assuming the cluster is relaxed, we then may integrate to give
\begin{equation}
n_p(E_p,r) = {Q_p(E_p)\over D(E_p)}{1\over 4 \pi r} \;,
\end{equation}
where $Q_p(E_p) = Q_0 E_p^{-s_p}$, and $Q_0$ is normalized to synchrotron observations.

The resonant diffusion of protons in magnetic field fluctuations with a power spectrum $P(k)$ in
wave number $k$ is given by (Blasi \& Colafrancesco 1999)
\begin{equation}
D(p) = 1/3 c r_{L}(p){B^2\over\int^\infty_{1/r_L(p)} P(k) dk}\;,
\end{equation}
where $r_L(p) = pc/eB$ is the Larmor radius. In this paper we take fluctuations of the
magnetic field to have a Kolmogorov spectrum,
\begin{equation}
P(k) = P(k_0) \bigg({k\over k_0} \bigg) ^{-5/3}\;,
\end{equation}
and assume the spectrum is normalized to
\begin{equation}
\int^\infty_{k_0} P(k) dk \approx B^2.
\end{equation}
Thus
\begin{equation}
P(k_0) = {2\over 3}{1\over k_0} B^2\;,
\end{equation}
and
\begin{equation}
D(p) = 1/3 c k_0^{-2/3} (eB)^{-1/3} E^{1/3}\;.
\end{equation}
A typical value for the smallest scale on which the magnetic field is homogeneous, 
$d_0 = 1/k_0$, is $> 20$ kpc, since the magnetic field is being stirred by the virial 
motion of galaxies within the cluster, for which $d_0 = (4 \pi/(3 N_{gal}))^{1/3} 
R_{cluster}$.

\section{Detection Counts}

In our model, the number of photons above $E_\gamma$ detected with GLAST,
writing $L_{43}$ as the luminosity in units of $10^{43}$ ergs s$^{-1}$, is (Berrington \& Dermer 2005)
\begin{equation}
N(>E_\gamma) \sim 35\, t_{yr} (E_\gamma/\hbox{\textrm{GeV}})^{-1.04}\;{L_{43}\over 0.3}\;,
\end{equation}
where we use a cluster evolution time $t_{yr} = 0.7$ Gyr, and $L_{43} = 0.1$ through $0.7$
coinciding with $s_p = 2.1$ through $2.5$. The detection significance
$\eta_\sigma$ for Coma, assuming it is a point-like source (i.e., the point spread function for a given
photon energy is larger than the $18^\prime$ angular extent of Coma) is (Berrington \& Dermer 2005)
\begin{equation}
\eta_\sigma \sim 5.4 \sqrt{t_{yr}} \bigg({L_{43}\over 0.3} \bigg) (E_\gamma/\textrm{GeV})^{0.1}\;.
\end{equation}
This yields a $\eta_\sigma = 5 \sigma$ detection significance.

We may also easily construct an argument for why neutrinos \emph{may} be detectable.
Suppose the $\gamma$-ray flux detected from the $\gamma$-ray bright Abell clusters is around
\begin{equation}
\Phi_\gamma(1 \;\textrm{TeV}) \approx 10^{-11} \ \textrm{photons} \ \textrm{cm}^{-2} \ \textrm{s}^{-1}
\end{equation}
at a few TeV (see Fig 4). This implies a similar number of muon neutrinos. Following
Gaisser \& Stanev (1984), the flux of muons produced in deep-inelastic interactions
between neutrinos and ice is nearly
\begin{equation}
{d\Phi_\mu\over dE_\mu} \approx {N_{Av} \sigma\over \alpha} E_\nu \Phi_\nu \approx
10^{-20} \ \textrm{GeV}^{-1} \ \textrm{cm}^{-2} \ \textrm{s}^{-1}\;,
\end{equation}
with the deep-inelastic cross section $\sigma$, muon energy loss rate $\alpha$, and
Avogadro number $N_{Av}$. The muons leave the interaction with about 1/3rd the incoming neutrino
energy, so that
\begin{equation}
\Phi_\mu \approx \bar{E_\mu} {d\Phi_\mu\over dE_\mu} \approx 3 \times 10^{-18} \;\textrm{cm}^{-2} \ \textrm{s}^{-1}.
\end{equation}
The area of ICECUBE being $A\approx10^6$ m$^2$, this gives $N_{Ice} = 1
(t_{\rm obs}/{\rm year})$.
That is, the number of neutrino detection events per year is approximately
$\Phi_{\gamma}(1\;{\rm TeV})/10^{11}$
(see, e.g., Crocker, Melia, and Volkas 2000, 2002, 2005; Crocker, Fatuzzo, Jokipii et al.
2005).
%This is borne out in a previous prediction (Crocker et al 2005) of neutrino flux based on HESS
%detection of the galactic center.
%*** I'm not clear on why this has been commented out. All I'm trying to say is that the neutrino
%*** number prediction should be something like $N_\gamma(GeV)/10^{-11}$ year$^{-1}$. Since
%*** even the interested reader is unlikely to go beyond the back-of-the-envelope I'd like
%*** to present at least one reference confirming this is the case.
%*** And, comparing the 2005
%*** paper's neutrino rate to the gamma flux levels, in looks as though in the case of the galactic center
%*** we've got that relation \emph{precisely}.

More precisely, the number of neutrino detection events in a particular device is
\begin{equation}
N_{\rm year} = \int dE_\nu \int^{year} \ dt \ {\rm Area}[E_\nu,\theta(t)] \ \Phi(E_\nu) \times P_{\rm detect}(E_\nu) \ {\rm Attn}[E_\nu,\theta(t)]\;,
\end{equation}
which includes attenuation of the flux as it passes through Earth,
the effective muon area---the preceding
depending on the location of the source defined in terms of its angle from
the nadir $\theta$---and
the probability of neutrino interaction and subsequent detection.
For the ICECUBE detector, these quantities are specified in detail in the Preliminary Design Document
available from the ICECUBE website (http://icecube.wisc.edu/). See also
Ahrens et al. (2004).

Carrying this out gives an event rate of $0.3$ detection events per year from Coma via the
ICECUBE detector. More optimistic are the detection rates should the $\gamma$-ray bright 
Abell clusters be confirmed: these range between $30$ upcoming events per year from Abell
$1914$ in the most optimistic case of a $s_p=2.1$ spectral index, to $8\times10^{-2}$
events per year from an extremely strong shock in Abell 85 (for the most pessimistic case).

Meanwhile, the prospect for using Auger to directly probe neutrino production via
p$\gamma$ collision processes in clusters is possible, but it would require very
favorable system parameters: even assuming
the weakest possible shock (and the most optimistic duty cycle of $15\%$),
yields only a few detection events per year, with Abell 1914 representing the
most promising source.

Table 2, summarizing these results, as calculated on the basis
of the results presented by Miele et al. (2006),
is given below; here, the fields separated by a semicolon indicate
decreasing proton spectra, from $s_p = 2.1$ to $2.5$, in steps of $0.1$.
At the $\sim$EeV energy scale of the p$\gamma$ neutrinos,
other experiments may offer apertures superior to that of Auger.
In principle, the ANITA experiment (Barwick et al. 2006), in particular, has
a considerably larger neutrino aperture than Auger but, unfortunately,
restrictions on the positions of potential sources (which must be located
within a narrow band of declination $[-9^\circ,19^\circ]$ to be detectable)
exclude all the potential cluster neutrino source considered here.

\section{Detections from Known Radio Clusters}

Let us suppose that $\gamma$-ray detections from Abell clusters 85, 1914, and 1758
are due to $\pi^0$ decay. We may then use multiwavelength observations to set
a very solid prediction for their neutrino flux, depending only on the spectral
index of protons---and their maximum energy. These are shown in Fig. 4 for a 
range of proton spectral indices consistent with Fermi acceleration and the 
synchrotron observations. 

The spectral slope of radio emission for Abell clusters 85, 1914, and 1758 are $>1.5$,
1.19, and 1.13, respectively (Ensslin et al. 1998).
Abell 85's value of $\alpha = 1.5$ would require a proton spectral index above that which could be produced
by shock acceleration. This could indicate either variability in measurements made years apart, or
the influence of several bright radio galaxies within the cluster. Here we take the maximum
value reasonable for shock acceleration, $s_p = 2.5$. The other values indicate proton
injection spectra near $s_p = 2.3$, matching the cooled spectrum of Coma (see Eqs. 13-15).

Of the three, Abell 85 appears to be the only cluster (Durret et al. 2005) for 
which temperature maps exist to create a radial gas profile, although Chandra 
detections of 1914 do exist (Govoni 2005). For Abell 85 the core gas density is
$7.7 \times 10^{-3}$ cm$^{-3}$, while in the other two clusters we assume it is 
the same as in Coma ($n_e = 3 \times 10^{-3}$ cm$^{-3}$). The flux of particles 
for each cluster scales as its relative redshift to Coma, while the active volume 
scales as the largest linear scale (LLS). In every case we take a nominal $B=1 \mu$G, 
for lack of better knowledge. This results in a strikingly similar value for the 
CR proton number density required to fit the radio flux (Fig. 3): in every case 
(including Coma) it is within a factor of two of $n_{CR} = 10^{-8}$ cm$^{-3}$, where
$n(p) = n_{CR} p^{-s_p}$. This is very near energy equipartition with the background 
thermal electrons. This balance in energy between the two species may be just
a coincidence, but more likely it represents a dynamic coupling between the protons
and the ambient electrons, such that an effective energy transfer occurs between
the two (see Wolfe and Melia 2008).

In each case,
$\gamma$-rays from $\pi^0$-decay dominate those from nonthermal bremsstrahlung by the synchrotron-emitting
electrons. It is therefore impossible to describe both the Coma cluster and these $\gamma$-ray bright clusters
within a simple secondary model, since Coma should then be a copious source of $\gamma$-rays. The situation
is not improved if one assumes nonthermal bremsstrahlung from the electrons produces the $\gamma$-rays,
since this would (via inverse Compton scattering) produce an X-ray excess not observed in Abell 1785 and 1914
(Abell 85, however, displays such an excess).

De Marco et al. (2006) have advanced the notion that photopion production between cosmic-ray protons
and infrared light emitted by galaxies within the cluster may dominate over proton-CMB interactions.
We agree with this general result, but find some difference in the details.
Again, pion production from p$\gamma$ interactions is characterized by a hard cut-off at the
threshold energy, followed by a spectral index one greater than the proton index, $s_\pi = s_p+1$.
The three-body decay then gives an extra degree of freedom, allowing neutrinos to be produced at
a constant emissivity until threshold. The De Marco et al. (2006) (their Fig. 5) and our own 
calculation include all of these features, except the location of threshold.
The energy of the galactic IR emission
under consideration (Lagache et al. 2003) is peaked a decade above the CMB peak; with the
threshold for $\gamma + p \rightarrow p + \pi$ being
\begin{equation}
E_{thres} = { (m_p + m_{\pi})^2 -m_p^2\over 4E_\gamma}\;,
\end{equation}
the Wein peak at $1.2 \times 10^{-3}$ eV gives a CMB threshold of $5.8 \times 10^{19}$ eV, while
the IR peak at $1.03 \times 10^{-2}$ eV gives a threshold of $6.6 \times 10^{18}$ eV (later
line emissions are suppressed by the steepness of the proton distribution).  Yet the threshold
of the De Marco et al. (2006) calculation is $\sim 10^{16}$ eV. This shift in the emission peak
means neutrino production begins at a point in the proton distribution some 100 eV in energy above
that given in De Marco et al. (2006), which with $n(p) \sim E_p^{-s_p}$ corresponds to a factor of 
$10^{-4}$ for the neutrino flux.

Finally, we comment on the possibility of calculating \emph{diffuse} cluster $\gamma$-ray and neutrino fluxes,
using the Coma cluster as a template. In this case it is helpful to look at those radio halo/relic
clusters which are known to exist, assuming that bright sources of $\gamma$-rays or neutrinos will
have first been observed in one of the radio surveys.
If cosmic-ray production is dominated by cluster mergers, the number of CR protons,
and thus the radio luminosity (as well as neutrino and $\gamma$-ray production),
should roughly scale as a cluster's mass, i.e., $L_p \sim GM^2/Rt_{\rm merge}
\sim (M/M_{\rm Coma})^{5/3}$ (De Marco et al. 2006). As a given cluster's mass is virially
connected to its temperature, we may predict the flux of cosmic rays and
neutrinos as a function of that cluster's relative temperature and
distance with respect to Coma.
In this case, $M \sim 5T R_v/G (k/\mu m_p)$, where we take the virial radius as scaling
with the largest linear size of a cluster. The $M-T$ relation has also been derived
from specific Chandra observations (Kotov \& Vikhlinin 2006) as
$M = A (T/5)^\beta \times 10^{14} M_\odot$ with
$A = 1.7$ and $\beta = 1.5$.

Should the population of radio halo/relics be powered by gravitational infall at the cluster's current mass,
we expect radio power to scale with the $5/3$rds power of cluster mass---but it does not
(Fig. 5). Using the Chandra $M-T$ relation gives a correlation between $(M/M_{\rm Coma})^{5/3}$
and $L_{\rm radio}$ of 0.46; the virial value improves this to 0.61. In no case is the relatively larger
proton density of a particular cluster greater than $(z_{\rm Coma}/z)^2$, meaning that none of
the known radio clusters could contribute a neutrino or $\gamma$-ray flux above that of Coma's.
The most concrete estimation for the diffuse flux---the weighted sum (by radio luminosity)
of the 30 or so known radio clusters---is approximately $9/4$ths that of Coma alone.
Of these, $\gamma$-ray bright double clusters Abell 1758 and Abell 1914 (Fig. 5, square
and diamond, respectively) are among the most dominant. The entire discussion 
is moot if the identification of Abell 85, 1758, and 1914 as $\gamma$-ray
bright clusters is confirmed, in which case these three clusters alone dominate the diffuse flux
despite their relatively low mass; in this case the importance of ongoing, violent mergers is emphasized.
A generalization which would allow a solid diffuse flux to be calculated appears to depend on a present
lack of knowledge about the basic mechanisms of cluster halos.

\section{Conclusions}

Observations of both $\gamma$-rays ($\gamma_{pp}$) and neutrinos (both $\nu_{pp}$ and $\nu_{p\gamma}$)
in clusters provide answers to a series of yes-or-no questions. If both 
$\gamma_{pp}$ and $\nu_{pp}$ are observed from a cluster, then proton-proton induced 
cascades are certainly at work and their contribution to the radio halo can be evaluated---as well
as potential models for the creation of viable proton populations with total energies $>10^{61}$ erg.
If $\gamma_{pp}$ is detectable by either EGRET or GLAST and $\nu_{pp}$ is not observed,
then $\gamma_{pp}$ is not likely to be caused by secondary decay (this is not a firm `no' as the
source in question may occupy an unfavorable position in the night sky, or the proton population
may experience a coincidental cutoff just after 100 MeV) and new models would need to be considered.
An Auger $\nu_{p\gamma}$ detection would confirm our expectation that very energetic cosmic-ray
photons interact with the CMB to produce debris particles, thereby losing their energy.
This would add some support to the view that cosmic ray events above $6 \times 10^{19}$ eV
cannot be hadronic if the incident particles originate from beyond the GZK limit.
Finally, if $\nu_{pp}$ is observed without $\nu_{p\gamma}$, we may assume that the spectral index of protons
is steeper than $2.1$, or that the cluster is incapable of accelerating protons to this energy.

The combination of X-ray, radio, $\gamma$-ray, and neutrino observations on galaxy clusters
would give a vital new perspective on the mechanisms producing high-energy particles in clusters.
A simultaneous detection of $\gamma$-rays and neutrinos from pp scatterings in a cluster would 
confirm conclusively that secondary production is at work, and provide a new probe for 
cosmic-ray production in the MeV and TeV ranges, as well as shock acceleration in clusters
and astrophysically in general. A detection (or upper limit) of p$\gamma$ neutrinos, 
meanwhile, would demonstrate whether clusters are capable of accelerating protons to ultra-high 
energies, and, indeed, how particle physics tested in colliders evolves towards these energies.

A confirmation via GLAST of Abell 85, 1758, or 1914, as $\gamma$-ray bright
would confirm them as strong neutrino sources, potentially detectable in
the ICECUBE and, perhaps, Auger experiments. If a nonthermal proton population 
approaching energy equipartition is demonstrated, these then form a dominant 
source of pressure in clusters, affecting structure and evolution. Of course,
the fact that neutrino detections are only possible near equipartition means
that there may be only a narrow window of opportunity for actually measuring
a neutrino flux, or maybe none at all. But we find it promising that an 
interesting prediction may be made even without requiring a super-equipartition
proton population. And, anyway, a near-equipartition situation between the
protons and the background electrons is suggested by a dynamic coupling between
these two populations, as discussed in Wolfe and Melia (2008). Taken together 
with evidence of ongoing mergers, these observations would give new insight 
into the period of cluster formation, beginning some $>9$ billion years ago, 
and what role mergers play in cosmic-ray production.

\acknowledgements

This research was supported by NSF grant AST-0402502 at the University of Arizona. This
work was supported in part by the Australian Research Council.
BW and FM are grateful for the hospitality of the Universities of Melbourne and Canterbury, 
where a portion of this work was carried out.

\newpage
%\begin{landscape}
\begin{table}[h1]
\caption{Properties of Clusters Containing Halo and/or Relic Candidates\label{tbl-1}}
\scriptsize
\begin{tabular}{lccccccccccc}
\tableline\tableline\\
Name 	& z 	& RA 		& DEC 		& L$_X$	(0.1-2.4 keV) 		& T 	& S$_{327}$ 	& S$_{1400}$ & Radio Power & LLS & L$_\gamma$
\tabularnewline
         &  	& (h m s)	& ($^o$ $\prime$ $\prime\prime$) 	& ($10^{44}$ erg s$^{-1}$) & (keV)
& (mJy) & (mJy) & $10^{24}$ W Hz$^{-1}$ & kpc &$10^{-8}$ cm$^{-2}$ s$^{-1}$\\
\tableline
A665     &0.1818  &08 30 47.4      &65 51 14        &14.78           &8.3	&108	&16 	 &17.6 	&1900\\
A697     &0.282   &08 42 57.6      &36 21 59        &16.30           &10.5	&29	&7 	&11.2	 &920\\
A725     &0.0921  &09 01 10.1      &62 37 20        &0.80            &7.3	&76	&6 	&3.1	 &440\\
A773     &0.2170  &09 17 54.0      &51 42 58        &12.35           &9.2	&35	&8 	&7.9	 &1400\\
A786     &0.1241  &09 28 49.7      &74 47 55        &1.53            &10.8	&319	&104 	 &21.9	&1400\\
A796     &0.1475  &09 28 00.0      &60 23 00        &1.38            &6.3 	&53	&8 	&5.6	 &1200\\
A1758    &0.2800  &13 32 45.3      &50 32 53        &11.2            &7.2	&55	&11 	&3.6 	 &1300	&9.2\\
A1914    &0.1712  &14 26 02.2      &37 50 06        &17.93           &10.7 	&114	&20 	 &6.2 	&1500	&16.3\\
A2034    &0.1130  &15 10 11.7      &33 29 12        &6.86            &7.0	&44	&8 	&1.0	 &920\\
A2061    &0.0777  &15 21 17.0      &30 38 24        &3.92            &5.5 	&104	&19 	 &21.5	&920\\
A2218    &0.1710  &16 35 52.8      &66 12 59        &8.77            &6.7	&9	&1 	&16.1	 &340\\
A2219    &0.2281  &16 40 22.5      &46 42 22        &19.80           &11.2 	&19	&2 	&2.6	 &810\\
A2255    &0.0809  &17 12 45.1      &64 03 43        &5.68            &7.3	&360	&18 	 &2.9	&930\\
A2256    &0.0581  &17 04 02.4      &78 37 55        &6.99            &7.3 	&1165	&190 	 &1.3	&1450\\
A2319    &0.0555  &19 21 05.8      &43 57 50        &12.99           &9.9	&204	&32 	 &5.3	&580\\
A1656C   &0.0232  &12 57 26.6      &28 15 16        &7.26            &8.2 	&3081	&640 	 &11.5	&1500\\
A1367    &0.0215  &11 40 18        &20 18 01        &1.6	     &3.5	&	&182	\\
A13      &0.0943 &00 13 32.2 &219 30 03.6   &2.24	&4.3				&&34	&1.3	 &880\\
A2744    &0.3080 &00 14 16.1 &230 22 58.8   &22.05	&11.04				&&38	&15.5	 &1700\\
A85      &0.0555 &00 41 48.7 &209 19 04.8   &8.38	&6.2				&&46	&0.61	&480	 &12\\
A115     &0.1971 &00 55 59.8 &126 22 40.8   &14.57	&9.8				&&80	&1.3	 &1500\\
A401     &0.0739 &02 58 56.9 &113 34 22.8   &9.88	&7.8				&&25	&0.6	 &590\\
A520     &0.2030 &04 54 07.4 &102 55 12.0   &14.20	&8.33				&&38	&6.7	 &1080\\
A545     &0.1540 &05 32 23.3 &211 32 09.6   &9.29	&5.5				&&41	&4.4	 &1500\\
A548     &0.0424 &05 45 27.8 &225 54 21.6   &0.30	&2.4				&&50	&0.4	 &360\\
A1300    &0.3071 &11 31 54.9 &219 54 50.4   &23.40	&5				&&14	&5.7	&780\\
A1664    &0.1276 &13 03 44.2 &224 15 21.6   &5.36	&6.5				&&107	&7.5	 &1400\\
A2163    &0.2080 &16 15 49.4 &206 09 00     &37.50	&13.83				&&55	&10.2	 &1500\\
A2254    &0.1780 &17 17 46.8 &119 40 48.0   &7.19	&7.2				&&32	&4.4	 &1140\\
A2345    &0.1760 &21 26 58.6 &212 08 27.6   &9.93	&8.2				&&92	&12.3	 &1200\\
A2390    &0.2329 &21 53 36.7 &117 41 32.2   &21.25	&10.13				&&69	&9.2	 &1560\\
\tableline\tableline\\
\end{tabular}
\end{table}
%\end{landscape}

\clearpage

\begin{table}[h2]
\normalsize
\caption{Neutrino Detection Event Rates\label{tbl-2}}
\begin{tabular}{lccc}
\tableline\tableline\\
Name 		& IC upcoming (1/year) 	& IC downgoing (1/year)	& Auger (1/decade)\\
\tableline	
Abell 85	& --			& 0.08; 0.3; 1; 3; 13	& 7e-5; 9e-4; 1e-2; 2e-1; 2\\
Abell 1914	& 0.2; 0.8; 2; 8; 30	& --			& 9e-5; 1e-3; 2e-3; 2e-1; 3\\
Abell 1758	& 0.1; 0.4; 1; 5; 18	& --			& 5e-5; 7e-4; 9e-3; 1e-1; 2\\
Coma		& 0.3			& --			& 0.03\\
\end{tabular}
\end{table}

\clearpage

\begin{figure}
\centering
\plotone{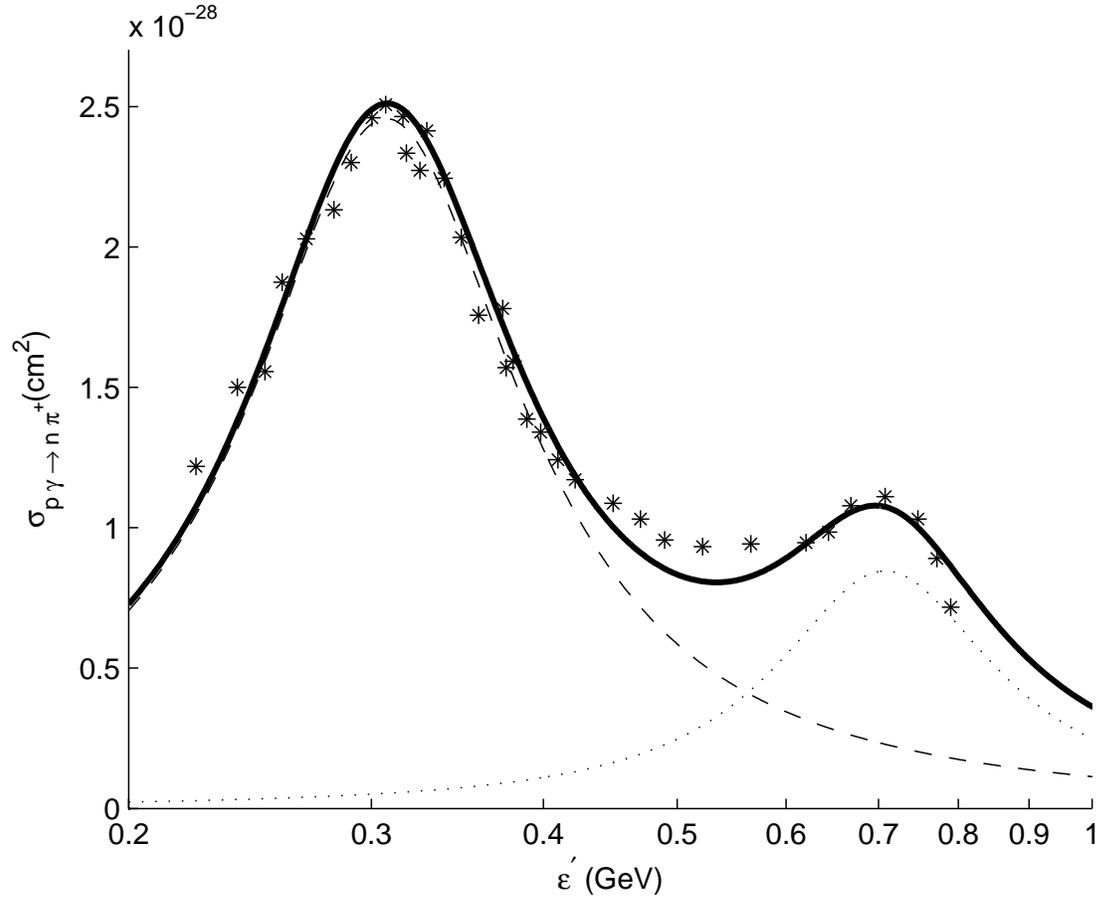}
\caption{
Cross-section for charged pion production via p$\gamma$ scattering, and its fit.
\emph{dashed}--the fit for the $\Delta$ resonance; \emph{dotted}--fit to the
$\epsilon^\prime \sim 0.7$ GeV resonance; \emph{solid}--overall fit;
\emph{stars} data from Mucke et al. (2000).
}
\end{figure}

%\begin{figure}
%\centering
%\plotone{f2.eps}
%\caption{
%Gamma-ray emission from Coma (thin solid) depends on the proton spectral index $s_p$, here displayed
%between $s_p = 2.1$ through $2.5$. It also depends on the radio emission, since a
%change in $s_p$ means renormalizing the CR proton density to fit these observations (thus
%$\gamma$ ray emission in the 70 MeV region increases with increasing spectral index).
%Also pictured are the detection limits for GLAST (1 year, dashed), EGRET (1 year, dash-dotted),
%VERITAS (thick solid), and MAGIC (5 $\sigma$, 50 hrs, dotted); as well as the nonthermal bremsstrahlung
%emission from the synchrotron-emitting electrons at $s_p = 2.1$ (solid shaded). $\pi^0$ decays will
%dominate $\gamma$ ray emission unless $B\ll 1 \mu$G.
%}
%\end{figure}

\clearpage

\begin{figure}
\centering
\plotone{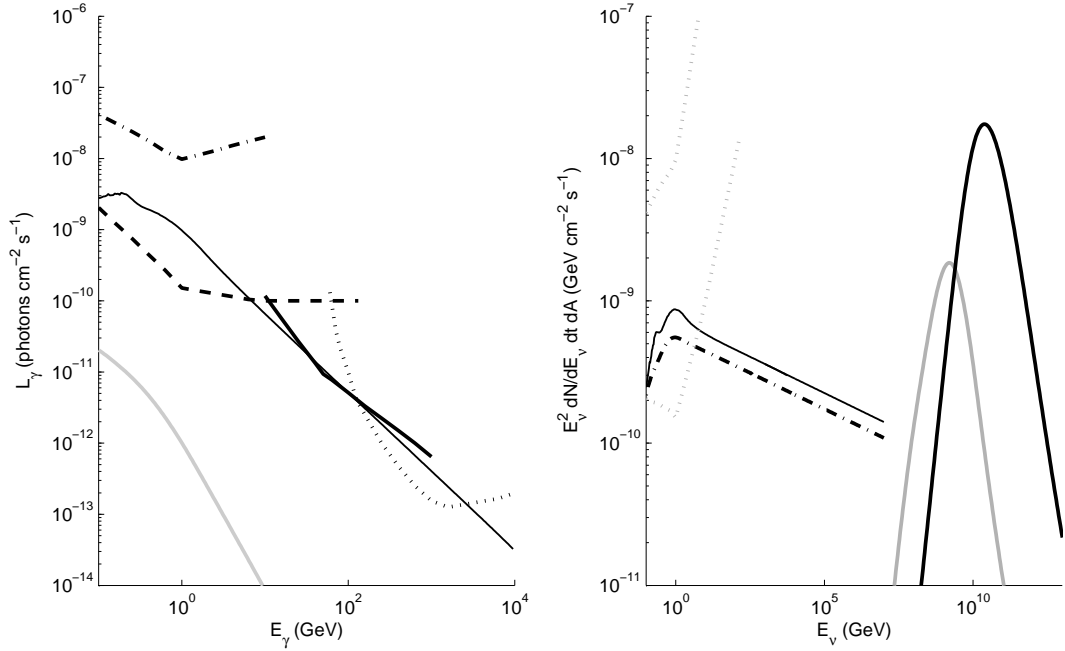}
\caption{
Gamma rays (left panel) and associated muon neutrinos (right panel) from the Coma cluster. In the left panel,
a flux of $L_{\gamma, 43} = 0.3$ is given by pion decay (thin solid),
which wholly dominates nonthermal bremsstrahlung from synchrotron-emitting electrons (solid shaded).
Also pictured are the detection limits for GLAST (1 year, dashed), EGRET (1 year, dash-dotted),
VERITAS (thick solid), and MAGIC (5 $\sigma$, 50 hrs, dotted). In the right panel we show Coma's associated
neutrino flux for pp scattering (dash-dotted), p$\gamma$ scattering with the IRB (solid shaded)
and that with the CMB (thick solid). The $\gamma$-ray detection limits (dotted) and flux (thin solid)
are shown for comparison.
}
\end{figure}

\clearpage

\begin{figure}
\centering
\plotone{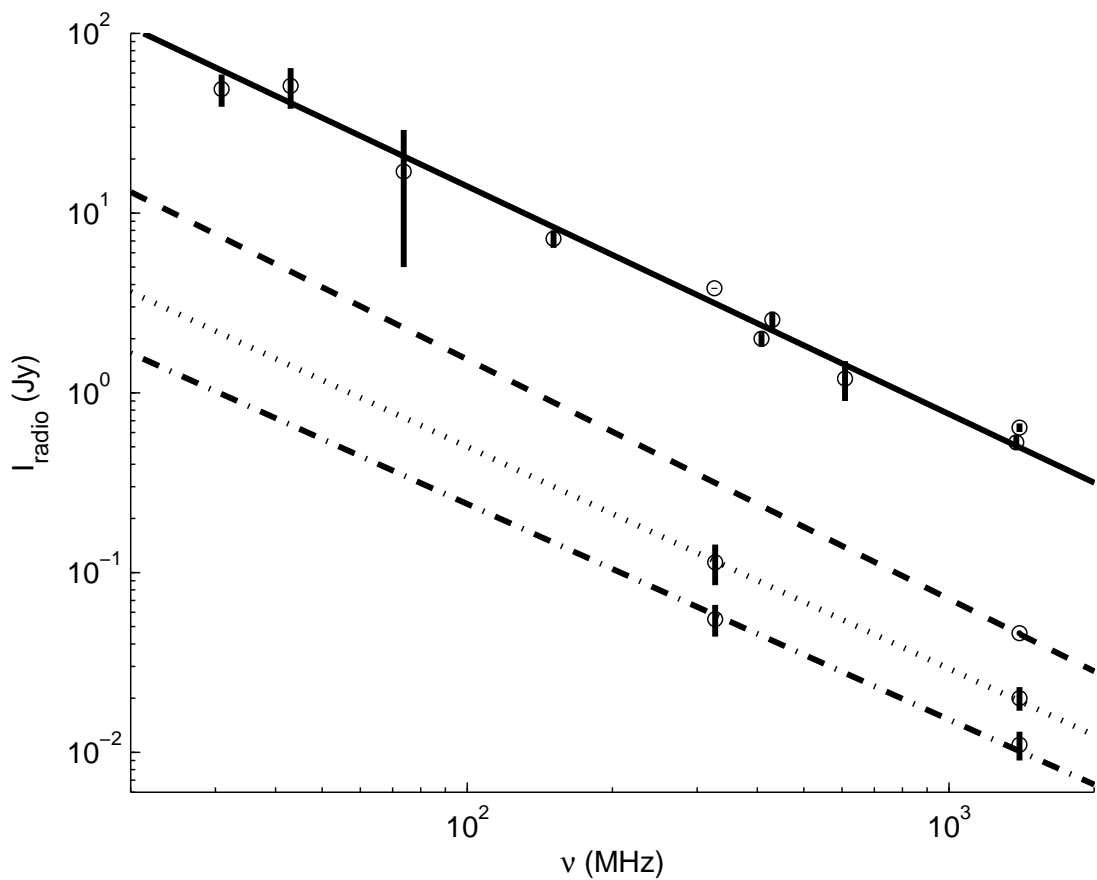}
\caption{
Radio emission from the four clusters sets the scale for the number and
index of cosmic rays.
\emph{solid}--Coma;
\emph{dashed}--Abell 1914;
\emph{dotted}--Abell 1758;
\emph{dash-dot}--Abell 85.
In each of the four clusters examined here, the number of cosmic-ray
protons is a few times $10^{-8}$ cm$^{-3}$, placing them just below energy equipartition
with the thermal background. The implied similarity of the three clusters' environments
(other than Coma)
means that, in each case, $\gamma$-ray emission is dominated by $\pi^0$ decay.
}
\end{figure}

\clearpage

\begin{figure}
\centering
\plotone{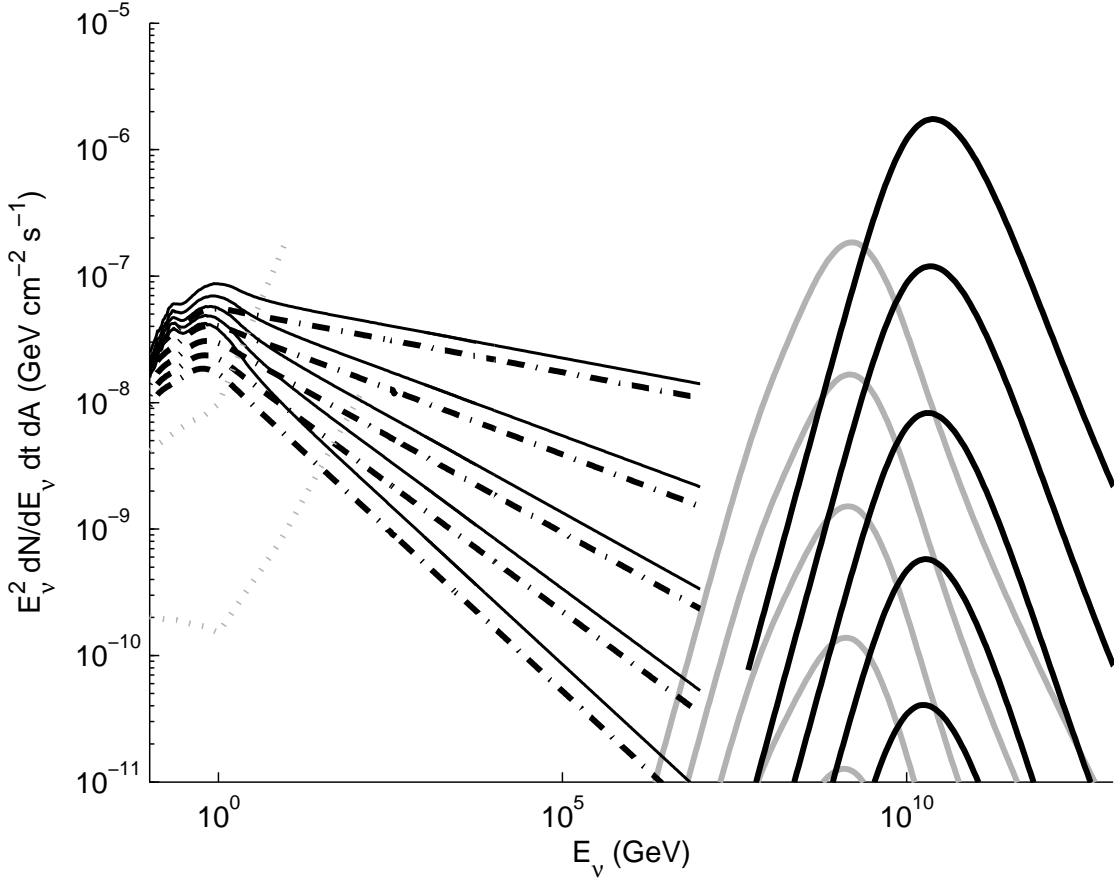}
\caption{
A positive $\sim$ GeV $\gamma$-ray detection directly implies a neutrino flux. In this case we display a
model for Abell 1914 which describes its $\gamma$-ray flux as a secondary model (solid, thin),
leaving the precise spectral index ($s_p= 2.1$ through $2.5$, in steps of $0.1$, moving downwards) 
as an unknown. The EGRET and GLAST limits (dotted) are again reproduced as a guide.
Neutrinos produced in concert with $\gamma$-ray emission from pp scattering (exclusively
$\nu_\mu$ produced via muon decay, dash-dotted)
dominate to several decades of TeV; those produced in p$\gamma$ interactions
with an infrared background (solid, shaded) peak at a few times $10^{18}$ eV
and dominate at a few times PeV; and p$\gamma$ interactions with the CMB (thick solid)
dominate above EeV.
}
\end{figure}

\clearpage

\begin{figure}
\centering
\plotone{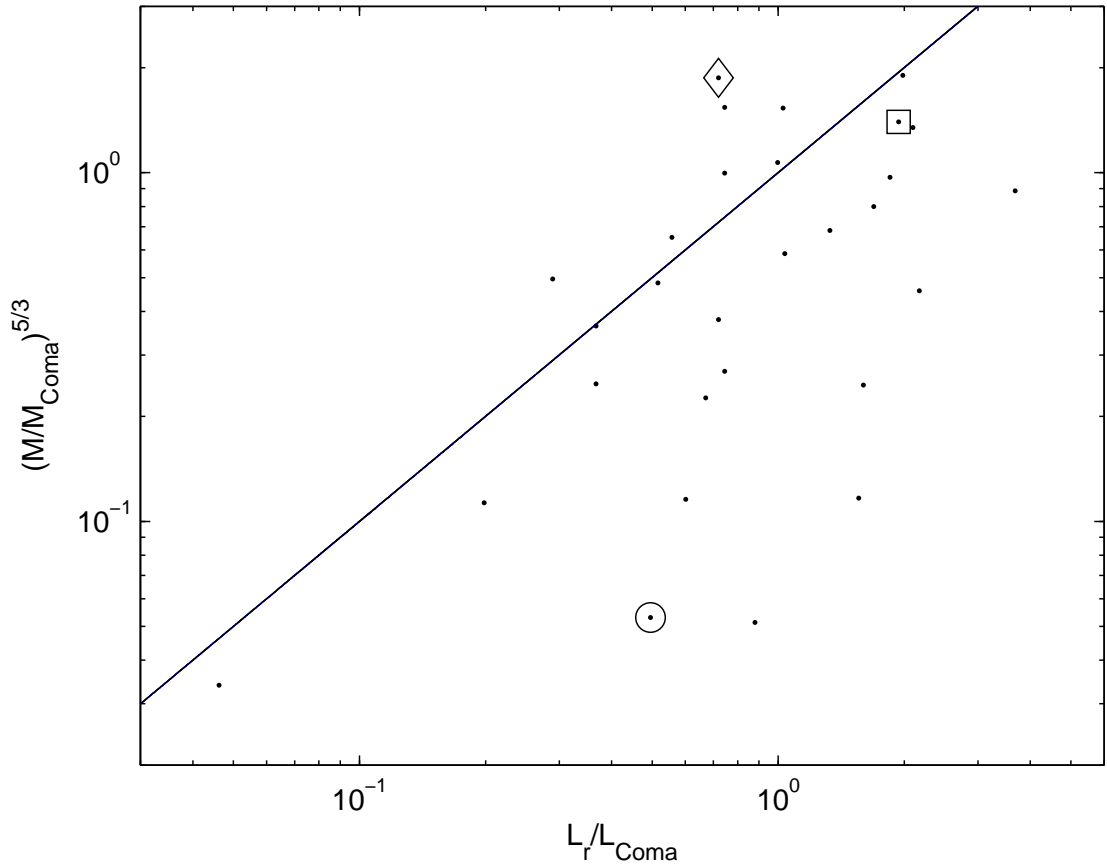}
\caption{
Radio luminosity of radio halo/relics does not scale as their mass. It is therefore difficult
to produce a template for calculating the total diffuse flux. The $\gamma$-ray bright radio halos
Abell 1758 (box) and 1914 (diamond) are relatively more massive than other clusters, which may
be why they are observed via EGRET.
}
\end{figure}

\end{document}